\documentstyle[11pt,aip]{article}

\def\eqref#1{Eq.~(\ref{#1})}

\def\figref#1{Fig.~\ref{#1}}

\begin{document}

\begin{center}
{\Large\bf A new fast method for}\\
{\Large\bf determining local properties}\\
{\Large\bf of striped patterns}\\
\ \\
{\bf David~A.\ Egolf$^\ast$}\\
{\bf Ilarion~V.\ Melnikov}\\
{\bf Eberhard Bodenschatz}\\
Laboratory of Atomic and Solid State Physics\\
Cornell University\\
Ithaca, New York, 14853 USA\\
$^\ast$ Also the Cornell Theory Center, Cornell University\\
February 6, 1997
\end{center}

\ \hrulefill\   

{
\bf  From the striped coats of zebras to the ripples in
windblown sand, the natural world abounds with locally banded patterns.  
Such patterns have been of great interest
throughout history, and, in the last twenty
years, scientists in a wide variety of fields have been studying
the patterns formed in well-controlled experiments that yield
enormous quantities of high-precision data.  These experiments involving
phenomena as diverse as chemical reactions in shallow layers
\cite{Ouyang91},
thermal convection in horizontal fluid layers \cite{Morris96}, 
periodically shaken layers of sand \cite{Melo95},
and the growth of slime mold colonies \cite{Lee96}
often display patterns that appear qualitatively
similar \cite{Cross93}.
Methods are needed to characterize in a reasonable amount of time
the differences and similarities in patterns
that develop in different systems, as well as in patterns formed 
in one system for different experimental conditions.
In this Letter, we introduce a novel, fast method for determining local
pattern properties such as wavenumber, orientation, and curvature
as a function of position for locally striped patterns.}

\figref{basic-pattern} shows a locally striped pattern
formed in a simulation of Rayleigh-B\'{e}nard convection,
in which a horizontal layer of fluid of lateral extent $L$
is confined between two parallel
plates with separation $d$ and a temperature 
difference $\Delta T = T_{\rm bottom} - T_{\rm top} > 0$.  For
$\Delta T$ larger than a critical
value~$\Delta T_c$, a convective
instability develops.  In most situations studied experimentally,
this instability leads
to the development of a striped convective roll pattern
with a horizontal local wavenumber $k \approx \pi/d$.
For systems of large aspect ratio~$\Gamma = L/d$ 
(i.e., many convective rolls), the patterns formed are often
spatially and temporally disordered but are still locally
striped as seen in
\figref{basic-pattern}.  The development of disordered locally periodic
patterns through instabilities
is common to a wide variety of large aspect
ratio experiments in biological, chemical, optical, and
fluid systems.

The use of local pattern properties to describe patterns
similar to that in \figref{basic-pattern} has already been explored
by several researchers.  Heutmaker and Gollub \cite{Heutmaker87}
determined the wavevector field and statistics such as the
``roll bending'' and ``roll obliqueness'' in both stationary and
time-dependent patterns in a circular Rayleigh-B\'{e}nard
convection cell.  They studied whether these quantities could be used
to understand the stability of convective patterns.  More recently,
Hu, et al \cite{Hu95} 
computed local wavenumbers and curvatures in
experimental pictures very similar to the one shown in
\figref{basic-pattern}.  Based on these local measurements,
they proposed order parameters to describe transitions in 
spatiotemporal chaos in Rayleigh-B\'{e}nard convection.
In a numerical study of a model of Rayleigh-B\'{e}nard
convection rotated about a vertical axis,
Cross, et al\cite{Cross94} computed
the local orientation of rolls 
to characterize domain structure.
Ouyang and Swinney \cite{Ouyang91} also
used the local orientation to
analyze patterns in a chemical reaction-diffusion system.
Gunaratne, et al \cite{Gunaratne95} have suggested an
invariant measure of the disorder of locally striped
patterns, and they have demonstrated its use on reaction-diffusion
patterns.
In a set of recent papers, Newell and
coworkers~\cite{Newell96,Passot94,Newell93def} 
have begun to use local wavenumbers
to study the behavior of phase-diffusion equations in the presence
of defects.

Although these previous studies clearly demonstrated the 
potential utility of local pattern properties, they were somewhat limited 
by the large amount of
time it took to process each snapshot of a complicated
time-dependent behavior.  The researchers used a variety of
methods including nonlinear least squares fits to small patches of
the pattern \cite{Gunaratne95}, automated \cite{Hu95} and
semi-automated \cite{Heutmaker87} determination of pattern
``skeletons'', and modern wavelet-based procedures
\cite{Ouyang91,Cross94,Newell96,Passot94,Newell93def}.
In this Letter we describe a new, fast method for calculating these
local properties based on ratios of simple partial derivatives.

For patterns that are locally striped, we can approximate
each field point $u(\vec{x})$ using:
\begin{equation}
u(\vec{x}) = A(\vec{x}) \cos(\phi(\vec{x})).
\label{u-approx}
\end{equation}
The local wavevector $\vec{k}$ is defined as:
\begin{equation}
\vec{k}(\vec{x}) \equiv \vec{\nabla} \phi(\vec{x}).
\end{equation}
Sufficiently far from defects and grain boundaries, we expect that
the variations in $A(\vec{x})$ are small compared to the variations in
$\phi(\vec{x})$.  In that case, the components of the wavevector
$\vec{k}$ can be found using simple partial derivatives.
For example,
\begin{equation}
k_x^2 = - \frac{\partial_x^2 u(\vec{x})}{u(\vec{x})},
\label{kx2-eq}
\end{equation}
where $k_x = \vec{k} \cdot \hat{x}$ and $\partial_x^2 \equiv 
\partial^2/\partial x^2$.  A similar expression can be written
for $k_y^2$.  However, two problems are immediately apparent.
The first is that when $u(\vec{x})$ is close to zero,
\eqref{kx2-eq} will be
very sensitive to small uncertainties in $u(\vec{x})$ due
to experimental or numerical noise.  This problem is easily
remedied by taking the ratio of the third partial derivative
to the first partial derivative for points where $u(\vec{x})$ is small.

The second problem is more subtle.  \eqref{kx2-eq} and the equivalent
equation for $k_y^2$ yield only the {\sl magnitudes} of the wavevector
components $k_x$ and $k_y$.  From these magnitudes, the vector can
be specified in any of four directions.  We note, however, that periodicity
in a real field is invariant under rotations by $\pi$, so
the wavevector $(k_x, k_y)$ is equivalent to the
wavevector $(-k_x, -k_y)$; i.e., the wavevector is actually a
{\sl director}, so we only need to determine
the relative signs of $k_x$ and $k_y$.  
We accomplish this by using combinations
of partial derivatives in the $x$-direction and the $y$-direction.
For example, if \eqref{kx2-eq} is used to determine $|k_x|$, the 
value of $k_y$ with its sign relative to $k_x$ is computed using:
\begin{equation}
k_y = k_x \frac{\partial_{xy} u(\vec{x})}{\partial_x^2 u(\vec{x})}.
\end{equation}
If $\partial_x^2 u(\vec{x})$ is close to zero, which occurs when
the wavedirector lies in a direction close to the $y$-axis,
$k_y$ is determined first and then $k_x$.

Our procedure for
determining the values of $k_x$ and $k_y$ at each point of the 
field $u(\vec{x})$ is summarized as follows.  
The global mean of the field, 
$<u(\vec{x})>_{\vec{x}}$,
is first subtracted from the field.  This result is then 
normalized so that the largest absolute value of the field is 1.
For each point $\vec{x}$, the value of the 
normalized field $u(\vec{x})$ and its partial
derivatives $\partial_x u(\vec{x})$, 
$\partial_y u(\vec{x})$, 
$\partial_x^2 u(\vec{x})$, and
$\partial_y^2 u(\vec{x})$ are used to choose
between the following four cases:
\begin{enumerate}
\item
$|u(\vec{x})| > \max(
|\partial_x u(\vec{x})|,
|\partial_y u(\vec{x})|)$ and
$|\partial_x^2 u(\vec{x})| > 
|\partial_y^2 u(\vec{x})|$:
\begin{eqnarray}
k_x & = & \sqrt{\left|
\frac{\partial_x^2 u(\vec{x})}{u(\vec{x})}\right|} \label{kx1-eq}\\
k_y & = & k_x
\frac{\partial_{xy} u(\vec{x})}
{\partial_x^2 u(\vec{x})}
\end{eqnarray}

\item
$|u(\vec{x})| > \max(
|\partial_x u(\vec{x})|,
|\partial_y u(\vec{x})|)$ and
$|\partial_x^2 u(\vec{x})| \leq
|\partial_y^2 u(\vec{x})|$:
\begin{eqnarray}
k_y & = & \sqrt{\left|
\frac{\partial_y^2 u(\vec{x})}{u(\vec{x})}\right|} \label{ky2-eq}\\
k_x & = & k_y
\frac{\partial_{xy} u(\vec{x})}
{\partial_y^2 u(\vec{x})}
\end{eqnarray}

\item
$|u(\vec{x})| \leq \max(
|\partial_x u(\vec{x})|,
|\partial_y u(\vec{x})|)$ and
$|\partial_x u(\vec{x})| > 
|\partial_y u(\vec{x})|$:
\begin{eqnarray}
k_x & = & \sqrt{\left|
\frac{\partial_x^3 u(\vec{x})}{
\partial_x u(\vec{x})}\right|}\\
k_y & = & k_x
\frac{\partial_y u(\vec{x})}
{\partial_x u(\vec{x})}
\end{eqnarray}

\item
$|u(\vec{x})| \leq \max(
|\partial_x u(\vec{x})|,
|\partial_y u(\vec{x})|)$ and
$|\partial_x u(\vec{x})| \leq
|\partial_y u(\vec{x})|$:
\begin{eqnarray}
k_y & = & \sqrt{\left|
\frac{\partial_y^3 u(\vec{x})}{
\partial_y u(\vec{x})}\right|}\\
k_x & = & k_y
\frac{\partial_x u(\vec{x})}
{\partial_y u(\vec{x})} \label{last-eq}
\end{eqnarray}

\end{enumerate}
If the value of $k_x$ is negative, the signs of both $k_x$ and $k_y$ are
changed to ensure that $k_x \geq 0$, while $k_y$ contains the
information about the orientation of the wavedirector.

Because this method is local, the effects of noise, of higher harmonics,
and of amplitude variations
neglected in \eqref{u-approx} are often noticeable.  We reduce these
effects by smoothing the $k_x$ and $k_y$ fields over small regions.
Although more sophisticated smoothing methods are possible, we have
successfully employed a simple method in which the $k_x$ (or $k_y$)
value is replaced by the average of the 
$k_x$ (or $k_y$) values within a small
square region of the field points.  We have generally found it sufficient
to smooth over
a region of size $\lambda^2$, where $\lambda$ is approximately the
average wavelength in the pattern.  The particular choice of $\lambda^2$
removes most of the contributions from
higher harmonics and amplitude variations.
Further smoothing can be accomplished
by repeating the procedure multiple times as necessary.
The size of the smoothing region and the number of smoothings can be 
easily determined empirically since local higher harmonics will 
appear as obvious modulations to the wavenumber field.

As we noted above, the wavevector field is actually a director
field, so care must be taken when averaging the 
neighboring $k_x$ ($k_y$)
values since $\vec{k}$ is equivalent to $-\vec{k}$.
To determine whether to include $\vec{k}$ or $-\vec{k}$
in the average, we take the
dot product of the wavevector at the point of interest,
$\vec{k_0}$, and the wavevector of the point to be
included in the
average, $\vec{k}$.  If $\vec{k_0} \cdot \vec{k} \geq 0$, the
angle between $\vec{k_0}$ and $\vec{k}$ is less than $\pi/2$ so
we average in $k_x$ ($k_y$); however, if $\vec{k_0} \cdot \vec{k} < 0$,
we average in $-k_x$ ($-k_y$).

Since the local wavenumber is undefined at defects and grain boundaries, our
procedure yields spurious values of $k_x$ and $k_y$ at these points.
In the averaging procedure we are careful to exclude values of
$k_x$ and $k_y$ for which $k^2 = k_x^2 + k_y^2$ is far outside the
range of expected values.  We have found that typically only values
within a distance
of about 1-2 field points around the defects need to be excluded from
the average.

Two other factors need to be considered when analyzing experimentally
generated data fields (as opposed to fields generated by simulating
partial differential equations).  Typically, experimental data have
a larger component of large-wavenumber noise than simulational data.
The effect of this noise can be lessened by applying a Fourier filter
to the entire data field before our procedure is employed.  This Fourier
filter should only be applied at wavenumbers well outside the expected
range of local wavenumbers.  Another potential problem
is that in some experiments the mean of the 
picture varies spatially, perhaps due to optical imaging effects.
One solution to this problem is to first determine the mean of the field
everywhere through averages over large regions and subtract this
spatially varying mean from the data field.  However, for fields in
which the variation of the mean is slow compared to typical local
wavelengths, a simpler solution is to replace the ratios
$\partial_x^2 u(\vec{x})/u(\vec{x})$ and $\partial_y^2 u(\vec{x})/u(\vec{x})$
in Eqs.~(\ref{kx1-eq}) and (\ref{ky2-eq}) with
$\partial_x^4 u(\vec{x})/\partial_x^2 u(\vec{x})$ and 
$\partial_y^4 u(\vec{x})/\partial_y^2 u(\vec{x})$, respectively.

A final consideration in this procedure is the best method for computing
the partial derivatives in Eqs.~(\ref{kx1-eq})--(\ref{last-eq}).
To obtain high-order derivatives, we perform the differentiation
in Fourier space.  Alternatively, the partial derivatives
could be obtained in coordinate space directly using high-order 
finite-difference schemes.  
We note that the calculation of the derivatives only consumes a
small fraction of the total time to evaluate a pattern.  The computation
is heavily dominated by the smoothing procedure and by the choice
of what derivatives to use at each field point.  If, for a given
application (such as locating disclinations), 
it is not necessary to determine the wavevector
field very precisely, little
or no smoothing is required and the computation time is reduced drastically.

Once the values of $k_x$ and $k_y$ are determined for each point of
a data field, local quantities such as the wavenumber, the
orientation of the periodicity, and the curvature
can be computed.
As a sample of the use of our procedure, we present four figures
showing these quantities for patterns
generated by simulations of the Boussinesq equations
describing Rayleigh-B\'{e}nard convection \cite{Pesch96}.  

The local orientation of the rolls, 
$\theta(\vec{x}) = \arctan(k_y(\vec{x})/k_x(\vec{x}))$,
at each point of the pattern 
in \figref{basic-pattern} is
shown in \figref{theta-pattern}.
This figure emphasizes the large regions of straight rolls often
found within the spiral defect chaos state of Rayleigh-B\'{e}nard
convection.
\figref{k-pattern} shows the local wavevector magnitude, 
$|\vec{k}(\vec{x})| = \sqrt{k_x^2(\vec{x}) + k_y^2(\vec{x})}$,
at each point of \figref{basic-pattern}.  The wavenumber is approximately
constant across the pattern with a mid-range value (green),
but small localized regions of high (red) and low (blue) wavenumber 
are interspersed throughout.  (These data, plus the value of the
roll curvature at each point, were obtained in less than 30~seconds
on a single RS/6000 POWER2 processor running at 66.7~MHz, and even less time
would be required at lower levels of smoothing.)

The region inside the small rectangle in \figref{basic-pattern}
is shown in greater detail in \figref{zoom-fig}(a).
\figref{zoom-fig}(b) and (c) show
the local roll orientation and local wavevector magnitude, 
respectively, for this pattern.
The rotation of the roll orientation about the centers of 
each of the two spirals is
shown clearly in \figref{zoom-fig}(b).  Also, along the left side and
along parts of the top and bottom portions of the figure are
regions of almost constant color indicating that the rolls are
approximately straight in those regions.  In \figref{zoom-fig}(c),
areas with compressed rolls (high wavenumber) 
are highlighted in red (e.g., the bottom
left corner), while defects and regions with low wavenumber are
shown as blue.

As an illustration of how these local pattern properties relate to the 
features of the pattern, 
\figref{single-defect}(a) shows a simple pattern with a single
dislocation moving up the page.
\figref{single-defect}(b) and (c) show the
magnitude of the wavevector and roll curvature 
$C = \vec{\nabla} \cdot \hat{k}$, respectively,
as a function of position for this pattern.  
The deformation of the pattern caused by the dislocation is clear
from this figure.  A ``wake'' of locally high wavenumber follows the
defect while small regions of low wavenumber form diagonally in front
of the defect.  The locality of the majority of the deformation is
also evidenced by the small regions of high curvature shown in
\figref{single-defect}(c).

We are currently studying the use of statistics based on 
the quantities described above
as order parameters for describing transitions between different 
spatiotemporal chaotic states.  In addition, we are looking at how the
spatial and temporal behavior of the local quantities are related to the
dynamics of the chaotic states.  Of particular interest is the
behavior of defects and disclinations, which can be located and
classified based on the local wavevector field.  These works will
be described in future publications.

We have presented a novel, fast, automated method for
determining local pattern properties as a function of spatial position 
for patterns that are locally striped.  Although we have 
emphasized our method's utility for characterizing spatiotemporal
chaotic states, we stress that our method is simple and general, and it
can be applied to locally striped patterns independent of their origin.
Because of its speed, this new method 
may allow for the calculation of new statistics to 
characterize disordered patterns in a wide variety of
scientific disciplines.

Correspondence should be directed to egolf@tc.cornell.edu.

We thank W. Pesch, A. Tschammer, and W.~B\"{a}uml 
for the use of their parallel codes to simulate
the Boussinesq equations.
This work was supported by the National
Science Foundation and by the Cornell Theory Center.

\newpage
\begin{thefigures}{99}

\figitem{basic-pattern}

The temperature field at the midplane of a simulation of the
Boussinesq equations with reduced Rayleigh number $\epsilon = 0.6$ and
Prandtl number $\sigma = 1.0$ in a box with square periodic lateral boundaries
with aspect ratio $\Gamma = 160 \times 160$ and 512 grid points
in each lateral direction.  Dark points indicate cold downflow; white
points indicate warm upflow.

\figitem{theta-pattern}

The local roll orientation 
$\theta(\vec{x}) = \arctan(k_y(\vec{x})/k_x(\vec{x}))$
in radians
for the field shown
in \figref{basic-pattern}.
Two smoothings, each with
a radius of 3 field points ($\approx \lambda/2$), are performed.

\figitem{k-pattern}

The local wavevector magnitude $|\vec{k}(\vec{x})|/k_c$ for the
field shown in \figref{basic-pattern}, where $k_c$ is the 
magnitude of the wavevector at the
onset of convection.  Two smoothings, each with
a radius of 3 field points ($\approx \lambda/2$), are performed.

\figitem{zoom-fig}
{\bf (a)}A detailed view of the pattern inside the small rectangle in
\figref{basic-pattern}, {\bf (b)} its local roll orientation, 
and {\bf (c)} its local wavevector magnitude $|\vec{k}(\vec{x})|/k_c$.  
The color schemes for (b) and (c) are the same as in
\figref{theta-pattern} and \figref{k-pattern}, respectively.

\figitem{single-defect}
{\bf (a)}
A small region ($\Gamma = 55 \times 48$) from
the temperature field at the midplane of a simulation of the
Boussinesq equations with reduced Rayleigh number $\epsilon = 0.55$ and
Prandtl number $\sigma = 1.1$ in a box with square periodic lateral boundaries
with aspect ratio $\Gamma = 201 \times 201$ and 512 grid points
in each lateral direction.  Dark points indicate cold downflow; white
points indicate warm upflow.
{\bf (b)} The local wavevector magnitude $|\vec{k}(\vec{x})|/k_c$
for the pattern shown in (a).  The colors range from blue (1.32) to
red (1.49).
{\bf (c)} The local roll curvature $C = \vec{\nabla} \cdot \hat{k}$
for the pattern shown in (a).  The colors range from
blue ($0.00 \times 10^{-3}\,k_c$) to red ($7.33 \times 10^{-3}\,k_c$).

\end{thefigures}

\end{document}